 \documentclass[final,5p,times,twocolumn,authoryear]{elsarticle}

\usepackage{amssymb}

\journal{Icarus}

\begin{document}

\begin{frontmatter}



\title{An Insolation Activated Dust Layer on Mars}


\author[a]{Caroline de Beule}
\ead{caroline.debeule@uni-due.de}
\author[a]{Gerhard Wurm}
\author[a]{Thorben Kelling}
\author[a]{Marc Koester}
\author[b]{Miroslav Kocifaj}

\address[a]{Faculty of Physics, Universit\"at Duisburg-Essen, Lotharstr. 1, D-47057 Duisburg, Germany}
\address[b]{Faculty of Mathematics, Physics and Informatics, Comenius University, Mlynsk$\acute{a}$ dolina, 842 48 Bratislava, Slovak Republic}

\begin{abstract}
The illuminated dusty surface of Mars acts like a gas pump. It is driven by thermal creep at low pressure within the soil. In the top soil layer this gas flow has to be sustained by a pressure gradient. This is equivalent to a lifting force on the dust grains. The top layer is therefore under tension which reduces the threshold wind speed for saltation. We carried out laboratory experiments to quantify the thickness of this activated layer. We use basalt with an average particle size of 67 $\mu$m. We find a depth of the active layer of 100 to 200 $\rm \mu m$. Scaled to Mars the activation will reduce threshold wind speeds for saltation by about 10\%.  
\end{abstract}


\end{frontmatter}

\section{Introduction} 
\label{introduction}

It is a long standing problem how to move particles on the martian surface. The most prominent mechanism is wind in analogy to transport of sand on Earth. Numerous work has been carried out on this in the past especially in wind tunnel experiments \citep{greeley1980, greeley1992, white1997}.

Recent images of the HiRISE camera onboard the Mars Reconnaissance Orbiter show that martian sand transport is still active. They find dunes which travel several meters in a few years \citep{bridges2012}. 
However, a problem encountered in the explanation of particle lift is that wind alone requires rather large  speed to initiate saltation. The pressure on Mars on average is only 6 mbar in contrast to 1000 mbar on Earth. This reduces the dynamic pressure of a gas flow strongly. A speed of 30 m/s is supposed to be necessary to pick up the most susceptible particles of 100 $\rm \mu m$ in size \citep{greeley1980}.
High tangential wind speeds in vortices might also mobilize particles. Obviously, dust devils bear witness of dust lifting. Visible dust devils come in a variety of sizes \citep{lorenz2009}. The largest ones might easily lift dust.
However, \citet{stanzel2008} and \citet{reiss2014} find wind speeds (tangential and transversal) which are not always large enough. Also, \citet{reiss2009} observed dust devil activity on Arsia Mons. This relates to an atmospheric pressure of only 2 mbar which requires still larger wind speeds.\\

There have been suggestions to support or ease particle lift one way or the other. The choice of particles to be picked up has been varied in wind tunnel experiments. As an example, the rolling of volcanic glass particles might reduce threshold speeds \citep{devet2014}. 
The pressure within dust devils is reduced compared to ambient conditions. It has been proposed that the traverse of such a pressure minimum might be sufficient to lift dust \citep{balme2006}. 
Last not least and connecting to the work presented here, \citet{wurm2006} found that illumination of a dust bed at low pressure provides a lift. This was applied to Mars by \citet{wurm2008}. Especially this latter effect is strongly depending on ambient pressure in a maybe non-intuitive way. Wind or gas drag and dynamic pressure decrease with decreasing pressure. The induced lifting force of an insolated surface can increase to lower pressure by orders of magnitude in strength. The force peaks around Knudsen numbers of $Kn \approx 1$, where $Kn$ is the ratio between the mean free path of the gas molecules and the size of a particle or pore within the dust bed. Hence, for micrometer dust particles insolation supported lift is not important on Earth but maximized on Mars.\\

The model discussed for this lifting force so far included photophoretic forces, solid state greenhouse effects and gas compression by thermal creep \citep{kocifaj2011, kelling2011a, debeule2013}. These are important on long timescales (hours) as current research is supporting (Koester et al. personal communication). However, here we consider gas flow through the dust bed and related pressure differences which was not included in the earlier models. This provides lift for a sample where illumination changes on short time scales of seconds or even fractions of seconds. The importance of this became obvious in microgravity experiments where \citet{debeule2014} observed an efficient gas flow through an illuminated dust bed directed upwards. The effect is tied to the temperature profile within the illuminated dust bed. 

The basic pumping is provided by thermal creep. For a constant temperature gradient pure thermal creep pumping does not require any pressure differences within the soil. However, the temperature profile along the top layer of an illuminated soil is rather flat as radiation is absorbed and thermally emitted. This layer cannot pump by thermal creep but has to keep up the thermal creep gas flow from below nevertheless. A pressure gradient is established close to the surface to do so. In accordance with Darcy's law the pressure increase below the surface transports the gas flow handed over by the Knudsen pump (thermal creep) below. Once set, this overpressure not only moves the gas but also acts on the dust particles. Particles within this top layer are under constant tension and can be ejected if cohesion and gravity can be overcome by any means. We call this an {\it insolation activated layer}.\\

We quantify the thickness of this layer here based on laboratory experiments. We illuminate a dust sample with sufficient light flux to compensate gravity. We then remove cohesion by short vibrations. This leads to a removal of the tension activated layer down to the pressure maximum and allows its thickness to be determined.

\section{Sub-surface pumping} \label{pumping}
\begin{figure}[ht]
\centering
\includegraphics[width= \columnwidth]{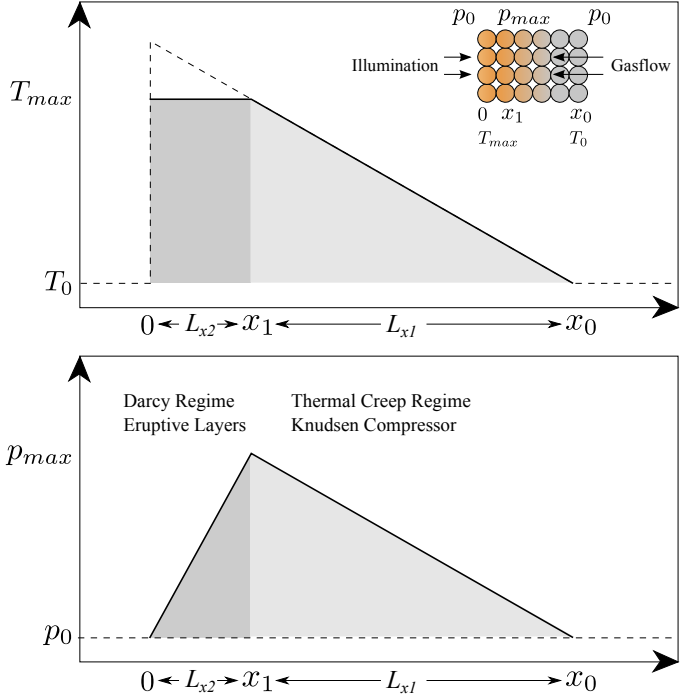}
\caption{\label{fig:principle} Principle for pressure distribution for given temperature profile and open geometry (ambient pressure the same on both sides of the dust sample). $L_{x2}$ marks the depth of the dust bed from $0$ to $x_1$ with constant temperature and $L_{x1}$ the part of the dust bed with a temperature gradient from $x_1$ to $x_0$.}
\end{figure}

We detail the light induced sub-surface pumping in this section. The principle is shown in Fig. \ref{fig:principle}. 

The light enters the dust bed and is absorbed. The heat is conducted further down into the dust bed. In addition, at the surface the dust bed can cool by thermal radiation. In consequence a temperature gradient is established starting a few particle layers within the dust bed (at depth $x_1$ in  Fig. \ref{fig:principle}) and is directed to deeper layers. 
Temperature gradients always lead to a thermal creep gas flow in a porous medium, where gas is transported along the particles' surface from cold to warm. 
This was first introduced by \citet{maxwell1879} as thermal transpiration where two gas reservoirs with different temperatures are connected by a small channel in a low pressure environment. Gas molecules creep along the channel wall from cold to warm. If the diameter of the channel is comparable to the mean free path of the gas molecules the pressure driven back flow can be smaller than the thermal creep flow. 
In an illuminated dust bed the gas molecules creep from cool layers deep within the dust bed ($x_0$ in  Fig. \ref{fig:principle}) upwards until the temperature levels off close to the surface at $x_1$. \\

If the temperature increase in Fig. \ref{fig:principle} would be linear from $x_0$ to the surface there would be no pressure differences. Every sub-layer would just transport the same amount of gas by thermal creep. However, if there is a top layer of constant temperature the gas molecules don't creep along these particles all the way to the surface, but only to $x_1$. The thermal creep gas flow leads to a concentration of molecules and the pressure is locally increased. \\
The increase of pressure below the surface at $x_1$ leads to a pressure driven gas flow through the top layer. The pressure adjusts itself to a value where gas flow through the top layer matches the incoming thermal creep gas flow from below. Both aspects, the temperature gradient driven thermal creep gas flow (Knudsen compression \citep{knudsen1909}) and the pressure driven gas flow (Darcy flow \citep{darcy1856}) are usually occuring in the same capillary.
The mass flow of the gas through capillaries was described by \citet{sone1990} and  \citet{muntz2002} as
\begin{equation}\label{eq:massflow}
\dot{M} = p_{avg} \frac{F A}{\sqrt{2 \frac{k_B}{\mu} T_{avg}}} \times \left(\frac{L_r}{L_x} \frac{\Delta T}{T_{avg}} Q_T - \frac{L_r}{L_x}\frac{\Delta p}{p_{avg}} Q_p \right)
\end{equation}
where $p_{avg}$ and $T_{avg}$ are the average pressure and temperature within the dust bed, $F$ is a factor giving the amount of capillaries within the surface area $A$, $k_B$ is the Boltzmann constant, $\mu$ the molecular mass of the gas, $L_r$ and $L_x$ are the radius and length of the capillaries and $\Delta T$ and $\Delta p$ are the temperature and pressure differences within the dust bed, respectively. The coefficients $Q_P$ and $Q_T$ depend on the Knudsen number and describe the pressure driven (back) flow and the flow by thermal creep, respectively. It has to be noted that the length of the capillaries $L_x$ is different in the thermal driven ($L_{x1}$, $|x_0 \rightarrow x_1|$) and the pressure driven ($L_{x2}$, $|0 \rightarrow x_1|$) part. \\

If the dust bed is heated two mass flows develop. The first one $\dot{M}_1$ is dominated by thermal creep (Knudsen pump) due to the temperature gradient
\begin{eqnarray}\label{eq:massflow1}
\dot{M}_1 = p_{avg} \frac{F A}{\sqrt{2 \frac{k_B}{\mu} T_{avg}}} \times \left(\frac{L_r}{L_{x1}} \frac{\Delta T_{L_{x1}}}{T_{avg}} Q_T - \frac{L_r}{L_{x1}}\frac{\Delta p_{L_{x1}}}{p_{avg}} Q_p \right) \; \mbox{with}\\
\frac{\Delta T_{L_{x1}}}{T_{avg}} Q_T > \frac{\Delta p_{L_{x1}}}{p_{avg}} Q_p 
\end{eqnarray} 
and the second one is driven by the pressure building up according to 
\begin{equation}\label{eq:massflow2}
\dot{M}_2 =  p_{avg} \frac{F A}{\sqrt{2 \frac{k_B}{\mu} T_{avg}}} \times \left(0 - \frac{L_r}{L_{x2}}\frac{\Delta p_{L_{x2}}}{p_{avg}} Q_p \right),
\end{equation} 
as the top layer temperature is flat or $\Delta T_{L_{x2}} = 0$.
It might be noted that the latter is also equivalent to a description by Darcy's law.
In total, the flow velocity is set by the Knudsen pump within the dust bed and the overpressure which maintains that flow also in the top layer activating it by putting tension on the dust particles. If the force caused by this pressure gradient overcomes gravity and cohesion, particles can be lifted from the dust bed's surface. \\ 

\begin{figure}[h]
\centering
\includegraphics[width=\columnwidth]{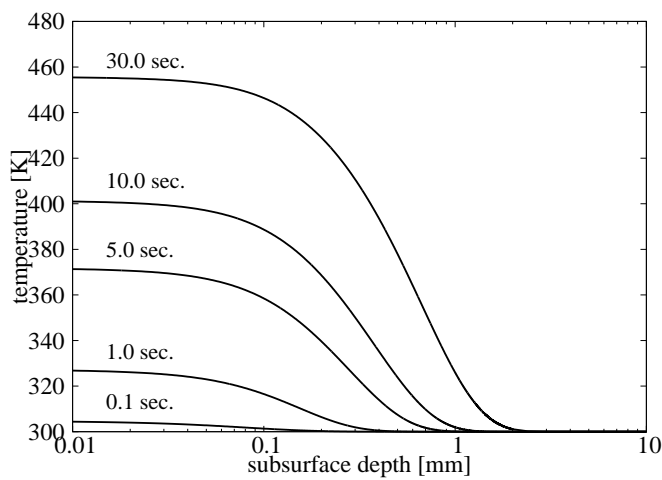}
    \caption{\label{fig:tempmax} Simulation of temperature profile within a dust bed consisting of 25 $\mu$m spheres. The dust bed has a thermal conductivity of 0.01 W K$^{-1}$m$^{-1}$. The light flux is 4 kW m$^{-2}$ and illumination times are on the order of seconds.}
\end{figure}

To quantify the temperature profile and the resulting flow velocities and pressure differences we modeled the insolation of a dust bed as described in earlier work by \citet{kocifaj2011}. Figure \ref{fig:tempmax} shows the numerical simulation of a temperature profile in a dust bed consisting of 25 $\mu$m (radius) spheres. 
The calculations were chosen to closely match the experimental setting: a composition of 25 $\mu$m sized (radius) spheres in a dust bed with a thermal conductivity of 0.01 W K$^{-1}$m$^{-1}$ (at $p = 10$ mbar \citep{presley1997}) and illumination with 4 kW m$^{-1}$ (see \citet{kocifaj2011} for details). 
The temperature starts to decrease beyond 100 $\rm \mu m$. Therefore, we expect that the activated layer would show a similar spatial scale.

\section{Experiment} \label{experiment}

\begin{figure}[ht]
\centering
\includegraphics[width= \columnwidth]{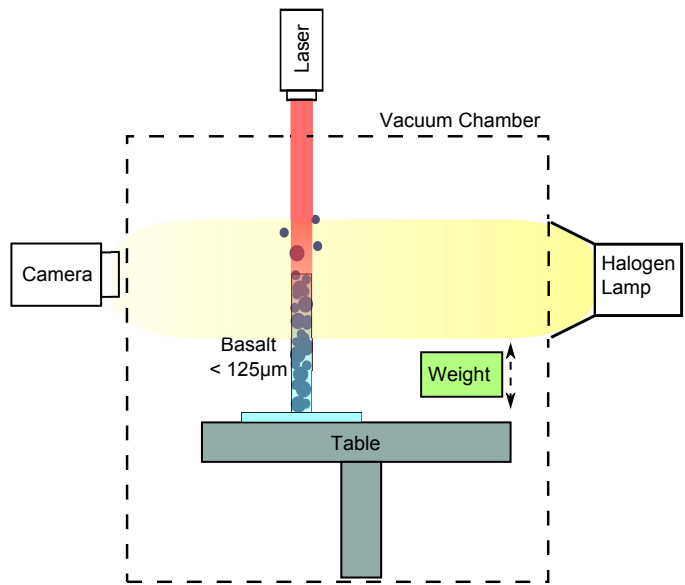}
\caption{\label{fig:setup} Schematic setup of the experiment: A dust sample is placed between glass plates of 2 mm distance on a table inside of a vacuum chamber with ambient pressure of 0.1, 1 and 10 mbar. The dust is illuminated by a red laser (655 nm) from above. A weight inside the chamber can be dropped upon the table, leading to a tension release within the dust. In addition some particles move upon the impact without being illuminated. Images are taken before and after the impact within and outside the laser beam. For the images outside the laser beam an additional light source is used to observe the surface with the camera.}
\end{figure}

\begin{figure}[h]
\centering
\includegraphics[width=\columnwidth]{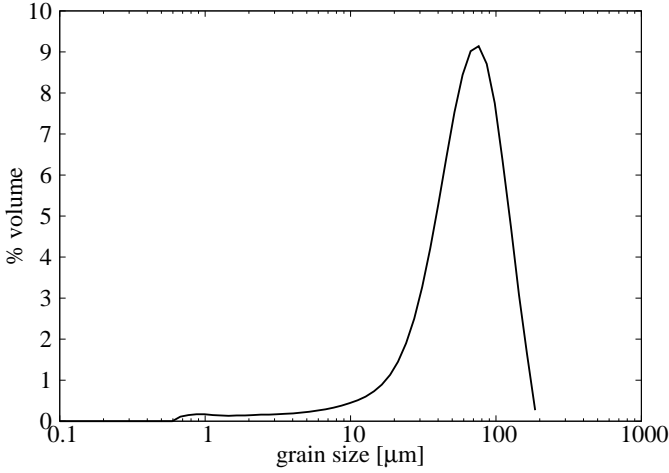}
    \caption{\label{fig:distribution} Volume distribution of the used sample.}
\end{figure}

The setup of the experiment is shown in Fig.\ref{fig:setup}. Basaltic dust with grain sizes ranging from 0 to 125 $\mu$m (average size of 67 $\mu$m (diameter) in a volume distribution, see Fig. \ref{fig:distribution}) is placed between two glass plates. It forms a thin dust bed with a width of 2 mm and 25 mm height.\\

About 0.5 mm of dust emerges above the plates, forming a smooth surface which can be observed by a camera and microscope optics. \\
In a distance of 3 cm to the glass slides a weight of 30 g is locked 8 mm above the experimental table by an electromagnet. If the magnet is turned off, the weight drops. This results in a short vibration of the dust sample, removing cohesion between the dust particles and -- if present -- releasing the tension due to the overpressure. 
This installation is placed within a vacuum chamber and experiments have been carried out at an ambient pressure of 0.1 mbar, 1 mbar and 10 mbar. A red laser (655 nm) illuminates a spot of 8 mm width of the dust sample for $\sim$ 20 seconds, providing a flux density of 4 kW m$^{-2}$. This flux is chosen to be close to the limit before a continuous particle ejection occurs. It provides the most tension to ensure that the whole activated layer is ejected on tension release.\\

\begin{figure}[ht]
\centering
\includegraphics[width=\columnwidth]{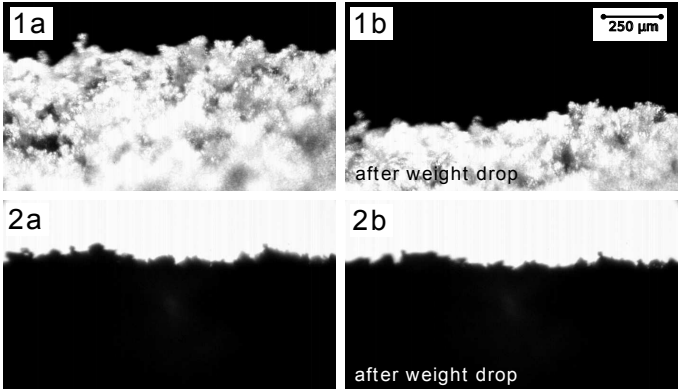}
    \caption{\label{fig:shape} Example of the dust bed within (1a) and outside (2a) the laser beam at 1 mbar ambient pressure. The images 1b and 2b show the results after tension release. The dust bed within the laser beam (1a, 1b) was illuminated by a red laser (680 nm) with a light intensity of 4 kW m$^{-2}$.}
\end{figure}

Images of the dust surface before and after the impact are taken. After each measurement the laser spot is moved further over the surface. After 5 measurements new dust is added to the sample. The images within the laser beam show a light surface and a black background (Fig. \ref{fig:shape}, 1a and 1b). In addition images outside the laser beam are taken before as well as after the impact to measure the particle loss without radiation. These are taken as bright field images with transmitted light to have a black surface on a bright background (Fig. \ref{fig:shape}, 2a and 2b).  \\

The surface lines before and after an ejection are traced. The particle loss is calculated by substracting the lines of the surface before and after the impact. For each image an average of the thickness of the removed layer with and without illumination is determined. A total of 40 to 50 averaged measurements is taken for each pressure.

\section{Model} \label{blowfish}

The mass flow of gas through capillaries can be calculated by Eq. (\ref{eq:massflow}).
We assume that two gas reservoirs are connected by a small capillary with a constant temperature gradient. For $Kn \approx 1$ the gas is driven by thermal creep and flows with a certain rate through the capillary until the rising pressure difference $\Delta p$ balances the thermal creep flow as a pressure driven flow through the top layer. 
There is a maximum $\Delta p$ at which no net gas flow is active ($\dot{M} = 0$, top layer as lid). From Eq. (\ref{eq:massflow1}) we get \citep{muntz2002}
\begin{equation}\label{eq:pmax}
\Delta p_{max} = p_{avg} \frac{\Delta T}{T_{avg}} \frac{Q_T}{Q_P}.
\end{equation}
We model the dust bed as follows: The mean radius of the dust particles is $r$ = 25 $\mu$m as used for the temperature profile in Fig.(\ref{fig:tempmax}). We consider this a suitable approximation for the dust sample of 33,5 $\mu$m radius (Fig.(\ref{fig:distribution})). We further approximate the radius of the capillaries by a pore size $L_r = 0.2  \times r$ \citep{jankowski2012}. The ambient pressure is $p = 10$ mbar (in one of the experimental sets). The illuminated area of the dust bed is $A =$ 2 mm $\times$ 8 mm, the Boltzmann constant $k_B = 1.38 \times 10^{-23}$ J K$^{-1}$. We assume a temperature difference $\Delta T = 100$ K over a length $L_{x1} = 1$ mm (see Fig. \ref{fig:tempmax}) and an average temperature of $T_{avg} = 273$ K $+ \Delta T / 2$ K within the heated part of the dust bed. According to \citet{sone1990}, the coefficients are $Q_T$ = 0.22 and $Q_P$ = 1.6. 
This results in $\Delta p_{max} = 0.4$ mbar.\\
In more detail the overpressure $\Delta p_{L_{x2}}$ equals $\Delta p_{L_{x1}}$ in Eq. (\ref{eq:massflow1}), and eventually, the gas velocity is the same in the temperature and the pressure driven part, as well as the mass flow. Solving the equation $\dot{M}_1 = -\dot{M}_2$ for the overpressure $\Delta p$, we get
\begin{equation} \label{eq:pressure}
\Delta p = \Delta p_{L_{x1}} = \Delta p_{L_{x2}} = \frac{L_{x2}}{L_{x1} + L_{x2}} p_{avg} \frac{\Delta T}{T_{avg}} \frac{Q_T}{Q_P}.
\end{equation}
Calculating this overpressure we take $L_{x1} = 1$ mm for the length of the capillaries with a temperature gradient and $L_{x2} = 0.1$ mm for the length of constant temperature  (see Fig. \ref{fig:tempmax}) we get an overpressure of $\Delta p = 0.05$ mbar. This is consistently lower that the estimate of the maximum pressure difference of $\Delta p = 0.4$ mbar.  
For an overpressure of about 0.05 mbar, the mass flow trough the pores is on the order of 10$^{-8}$ kg s$^{-1}$. 
The velocity of the gas flow can be calculated by dividing the mass flow by the illuminated area and the density of the gas at the given pressure $\rho_{gas} = p\; \mu / (T_{avg} R_g)$ with the molar mass of air $\mu = 28.96$ AMU and the molar gas constant $R_g$ = 8.3 J (K mol)$^{-1}$. At given parameters and $p = 10$ mbar, the gas flow is 11 cm s$^{-1}$. This is consistent with measurements of tracer particles in drop tower experiments by \citet{debeule2014}.\\
Further on, the force induced by the overpressure can be calculated for a particle column as
\begin{equation}\label{eq:force}
F_{p} = \Delta p  \; \sigma.
\end{equation}
With a particle cross section $\sigma = \pi r^2$ ($r = 25$ $\mu$m) we get $F = 6.7 \times 10^{-9}$ N. \\ 
The gravitational force which has to be overcome can be approximated by $F_G = m \; g \; L_{x2} (2r)^{-1}$ with $m = 4/3 \pi r^3 \; \rho$. With the density of basalt $\rho = 2890$ kg m$^{-3}$ we get $F_G = 3.7 \times 10^{-9}$ N, which is lower than the force caused by the overpressure by a factor of about 2. This is consistent with the fact that we adjusted the light flux to a value where no constant particle ejections occured, i.e. where gravity is compensated but not yet cohesion. As the overpressure is sufficient to compensate gravity of the top layer, only cohesion need to be overcome. As cohesion is removed by the impact the tension release leads to the ejection of the whole active particle layer.

\section{Results} \label{results}

\begin{figure}[ht]
\centering
\includegraphics[width=\columnwidth]{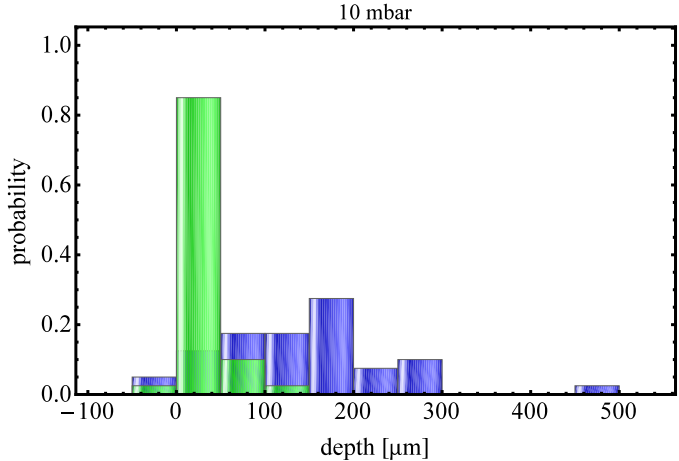}
    \caption{\label{fig:results10} Probability to find a certain depth of particle layer ejected at 10 mbar; blue bins are within the laser beam; green bins show the depth without radiation for comparison}
\end{figure}
\begin{figure}[ht]
\centering
\includegraphics[width=\columnwidth]{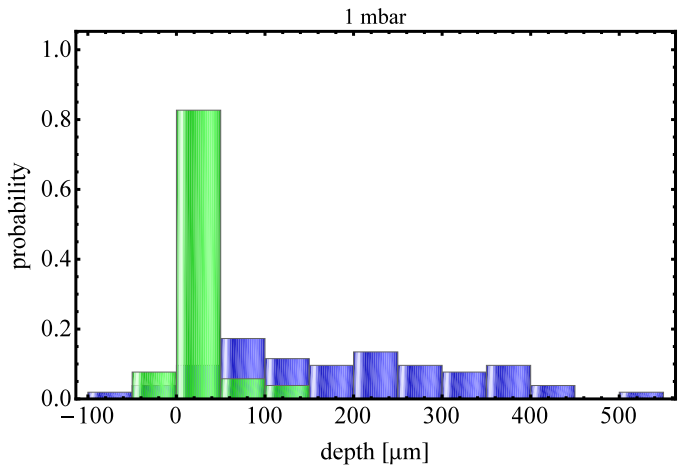}
    \caption{\label{fig:results1} Probability to find a certain depth of particle layer ejected at 1 mbar; blue bins are within the laser beam; green bins show the depth without radiation for comparison}
\end{figure}
\begin{figure}[ht]
\centering
\includegraphics[width=\columnwidth]{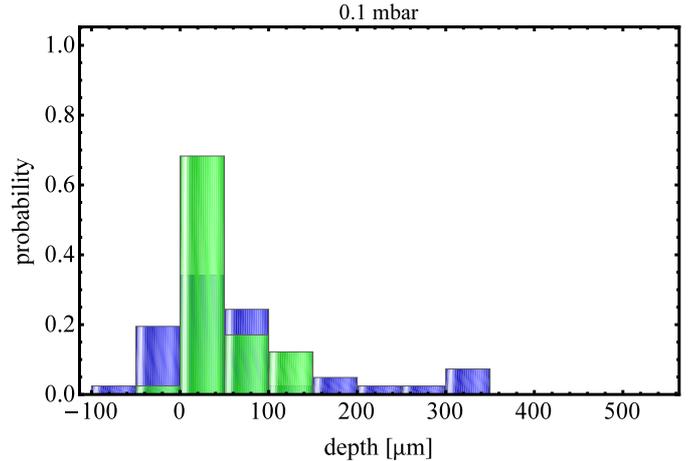}
    \caption{\label{fig:results2}  Probability to find a certain depth of particle layer ejected at 0.1 mbar; blue bins are within the laser beam; green bins show the depth without radiation for comparison}
\end{figure}

The measurements show a significant difference between the depth of particle loss within and outside the laser beam for each pressure value as seen in Fig. \ref{fig:results10}, Fig. \ref{fig:results1},  Fig. \ref{fig:results2} and in Tab. \ref{tab:meandepth}. This proves that there is an activated dust layer. The green bins in the figures show the probability of the depth to which particle loss occurs without light. The blue bins show the probability to find a certain layer depth within the laser beam. The negative depths refer to the fact that the surface can not be prepared perfectly smooth and therefore some aggregates can move during the tension loss, falling onto the measured spot and increasing the surface instead of decreasing it. Without light the measured values are strongly peaking around a depth within the dust bed of 50 $\mu$m. Within the illuminated area of the dust bed's surface the measured depth ranges from 50 up to 500 $\mu$m for 1 and 10 mbar, where the variation of the values seems somewhat higher for 1 mbar. For 10 mbar the layer size shows less variation and is more centered around 100 $\mu$m. The values for 0.1 mbar show no significant loss attributed to illumination. This is due to the decrease in absolute pressure and pressure difference. While the temperature profile is similar as in the other two cases it is not imprinting itself enough on an overpressure to dominate over Earth gravity. \\
The temperature profile for the relevant time scales in our experiment shown in Fig. \ref{fig:tempmax} is consistent with the experimental results. The flat temperature region has an extension of around 100 $\mu$m which is the same as the measured depths for 1 and 10 mbar.  Average values are given in Tab. \ref{tab:meandepth}.
\begin{table}[h]
	\begin{center}
	\begin{tabular}{ |c| c| c| p{3.7cm}| c| p{2.2cm}| }
		\hline
		 pressure  & mean thickness  & mean deviation  \\
		\hline
		0.1 mbar & 33 $\mu$m  & $\pm$ 32  $\mu$m \\
		1 mbar & 166 $\mu$m &  $\pm$ 57  $\mu$m\\
		10 mbar & 109 $\mu$m & $\pm$ 35  $\mu$m \\
		\hline
	\end{tabular}
\caption{The mean thickness of the active dust layers for the different pressures. Each data point is based on about 40 measurements and the mean thickness was calculated after subtracting the mean offset.}
\label{tab:meandepth}
\end{center}
\end{table}
\section{Application to Mars} \label{application}

\begin{figure}[h]
\centering
\includegraphics[width=\columnwidth]{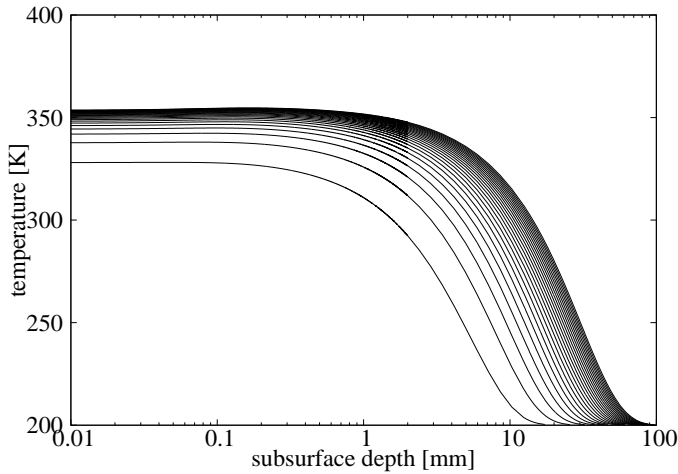}
    \caption{\label{fig:marstemp} same as fig. \ref{fig:tempmax} but for a light flux of 700 W m$^{-2}$,  average temperature of 200 K. The illumination times are 0.5 h, 1 h, 1.5 h$\ldots$ 12.5 h from bottom to top.}
\end{figure}

The laboratory setting was chosen to quantify the active layer depth in a convenient way. On Mars the light flux is and the average temperature are lower and the durations of insolation are longer. We therefore adjusted our simulation to scale the experimental results to martian conditions. We choose 700 W m$^{-2}$ for the solar insolation  and illuminate the dust bed in our simulation for up to 12.5 hours at an average temperature of T = 200 K.  The temperature profiles are shown in Fig. \ref{fig:marstemp}. The maximum temperatures are somewhat larger than temperatures found on Mars. This is due to the simplifications of the model but we consider the results to capture the essential trend. Additionally we consider CO$_2$ instead of air with a molecular mass of $m_{CO_2} =$ 44 $\mathrm{AMU}$, a geometric radius of $4.63 \times 10^{-10}$ m and a resulting cross section $\sigma = 1.6 \times 10^{-19}$ m$^2$. At p = 6 mbar with a particle density $n = p k_{B}^{-1} T^{-1}$ the mean free path of CO$_{2}$ is $\lambda = (\sqrt{2} n \sigma)^{-1} =$ 19 $\mu$m. Hence, the ratio between Q$_T$ and Q$_P$ is 1.68 (\citet{sone1990}). According to Fig. \ref{fig:marstemp} we expect typical temperature differences of $\Delta T = 150$ K for a length $L_ {x1} = 50$ mm (capillaries with a temperature gradient) and $L_ {x2} = 1$ mm for the activated layer of constant temperature. \\
Equation \ref{eq:pressure} gives a pressure difference of $\Delta p = 0.01$ mbar. This corresponds to a lifting force for a particle pile in the active layer of $F_P = 1.8 \times 10^{-9}$ N. Gravity on the same particles is $F_{G, Mars} = m \; 0.38 \; g  \; L_{x2} (2r)^{-1} = 1.4 \times 10^{-8}$ N. That means that the activation accounts for 13\% of gravity in a 1 mm thick layer of the soil. Variations in strength are expected for changing insolation and soil parameters.

\section{Conclusion} \label{discussion}
Our laboratory experiments showed that the illumination of a dust bed at low ambient pressure leads to an overpressure below the surface. This pressure provides a lifting force on particles in insolated dust beds. The pressure and thickness of this activated layer can well be described by a thermal creep model.  In the experiments we used 4 kW m$^{-2}$ and illuminated dust beds for about 20 s. This results in an activated layer of about 100 micrometer thickness. In the laboratory the overpressure at 1 mbar and 10 mbar is large enough to compensate (Earth) gravity. \\
At martian conditions the activated layer is supposedly deeper up to 1 mm. The supporting force can be on the 10\% level compared to gravity.  We expect that this mechanism supports other lifting mechanisms significantly, e.g. by decreasing the threshold wind speed for saltation by 10\%.\\

\section{Acknowledgements}

This project is supported by DLR Space Management with funds provided by the Federal Ministry of Economics and Technology (BMWi) under grant number DLR 50 WM 1242 and by the German Research Foundation (DFG) under grand number Ke 1897/1-1. We thank Ralph Lorenz and the anonymous referee for their thoughtful reviews. 





\end{document}